\documentstyle[epsfig,12pt]{article}
\def\beq{\begin{equation}}
\def\eeq{\end{equation}}
\def\to{\rightarrow}
\def\noi{\noindent}
\def\lsim{\raise0.3ex\hbox{$<$\kern-0.75em\raise-1.1ex\hbox{$\sim$}}}
\def\gsim{\raise0.3ex\hbox{$>$\kern-0.75em\raise-1.1ex\hbox{$\sim$}}}
\font\boldgreek=cmmib10
\mathchardef\mypsi="0920

\begin {document}
\baselineskip 24pt
\parindent=15pt
\vspace{0.5cm}
\begin{center}
{\large \bf   A QUANTITATIVE REANALYSIS OF CHARMONIUM}\\
{\large \bf SUPPRESSION IN NUCLEAR COLLISIONS}
\\ \vskip 1. truecm

{\bf N. Armesto and A. Capella} \\
{\it Laboratoire de Physique Th\'eorique et Hautes Energies,\\
Universit\'e de Paris XI, B\^atiment 211,
F-91405 Orsay Cedex, France}
\\
\end{center}
\vskip 2.0 truecm

\begin{abstract}
We present a quantitative description of $J/\psi$
and $\psi^\prime$ suppression in proton-nucleus
and nucleus-nucleus collisions at CERN energies. We use a conventional hadronic
framework
based on nuclear absorption plus final state interaction of the $J/\psi$
or $\psi^\prime$
with co-moving hadrons.
\end{abstract}

\vfill
\noi hep-ph/9705275 \\
\noi LPTHE Orsay 97/11 \\
\noi DESY 97-133 \\
\noi April 1998

\newpage
\pagestyle{plain}
Charmonium suppression due to Debye screening in a deconfined medium was
proposed in 1986 by Matsui and Satz \cite{1r} and found experimentally by the
NA38 Collaboration \cite{2r}. However, it was claimed very soon \cite{3r,4r,5r}
that this phenomenon, which is also present in $pA$
collisions, could be due to the absorption of the pre-resonant $c\bar{c}$ pair
in the colliding nuclei. It is nowadays known \cite{6r,7r} that the $J/\psi$
data on
$pA$
and
on $AB$ collisions with a light projectile can indeed be described by nuclear
absorption with an absorptive cross-section $\sigma_{abs} = 7.3 \pm 0.6$ mb.

Recently the NA50 Collaboration has found an anomalous $J/\psi$ suppression in
$PbPb$ collisions, i.e. a suppression which is substantially stronger than the
one obtained from nuclear absorption with the above value of $\sigma_{abs}$
\cite{7r,8r}. Two different interpretations of this anomalous suppression
have been
proposed in the literature. One is a scenario in which there is an extra
suppression of the $J/\psi$ due to interactions with co-moving
hadrons (co-movers)
\cite{9r,10r,11r}.
The other interpretation \cite{12r,13r,14r} assumes that when the
local energy density is larger than some critical value (taken to be
around the maximal
one reached in a central $SU$ collision), there is a discontinuity
in
the $J/\psi$ survival probability. (See also \cite{arm} for ideas based on
percolation of strings.)

In recent papers Kharzeev, Louren\c co, Nardi and Satz \cite{6r} and
Vogt \cite{vqgp}
have claimed that a quantitative analysis of the
data allows to conclude that the co-mover
model cannot describe the data, whereas
a quark-gluon plasma interpretation describes them well. In the present work we
re-examine
this point by reanalyzing all available data, using the final 1995 NA50
results
\cite{8r}, in the conventional co-mover
scenario mentioned above (see also \cite{ac97}). \par \vskip 5 truemm

\noi \underbar{\bf Nuclear absorption:}
We describe it in the probabilistic model of Ref. \cite{4r}.
Let us consider first
proton-nucleus collisions. In this model, the pre-resonant $c\bar{c}$ pair is
produced at some point $z$ inside the nucleus and scatters with nucleons
on its path
at $z^\prime > z$,
with an absorptive cross-section $\sigma_{abs}$. This produces a
change in the $A$ dependence of the $J/\psi$ inclusive cross-section. For
nucleus-nucleus collisions this change, at impact parameter $b$ and transverse
position $s$, is given \cite {4r} by

\beq
S^{abs}(b, s) = {[1 - \exp (- A \ T_A(s) \ \sigma_{abs}) ] [1 - \exp (- B\
T_B(b-s) \ \sigma_{abs})] \over \sigma_{abs}^2
\ AB \ T_A(s) \ T_B(b-s)} \ \ \ . \label{1e} \eeq

\noi Here $T_A$ and $T_B$ are the nuclear profile functions normalized
to unity, determined from a
standard Saxon-Woods density $\rho(r)=\rho_0/(1+\exp{[(r-R_A)/a]})$, with
$R_A=1.14\ A^{1/3}$ fm and $a=0.545$ fm \cite{nd}. $\sigma_{abs}$ is
the absorptive cross-section.
In the following we take $\sigma_{abs} = 7.3 \pm 0.6$
mb which gives the best fit to
the $pA$
data \cite{6r}.
Note that in Refs. \cite{7r,8r} a smaller value, $\sigma_{abs} = 6.2 \pm
0.7$ mb, has been obtained. This is due to the fact that the approximate
expression of nuclear absorption in Ref. \cite{5r} was used, instead of
Eq. (\ref{1e}).
Also note that $S^{abs} = 1$ for $\sigma_{abs} = 0$, so expression
(\ref{1e})
has the meaning of a survival probability of the $J/\psi$ due to nuclear
absorption.

Since we are aiming at a quantitative analysis it should be emphasized that the
probabilistic
formula, with its longitudinal ordering in $z$, can only be true in
the low energy limit. Therefore it is important to evaluate the uncertainty
resulting from using this formula at $\sqrt{s} \sim 20$ GeV. In a recent paper
\cite{15r}
the equivalent of Eq. (\ref{1e}) has been derived in a field theoretical
approach.
The obtained formula is valid at all energies and coincides exactly with
(\ref{1e})
in the low energy limit. It is amazing that the differences between the
results obtained
with this exact formula and the ones obtained from Eq. (\ref{1e})
are less than 1 $\%$. \par \vskip 5 truemm

\noi \underbar{\bf Absorption by co-moving hadrons:}
The survival probability of the $J/\psi$ due to absorption with co-moving
hadrons is given by (see \cite{6r,9r,10r} and references therein)

\beq
S^{co}(b, s) = \exp \left [- \sigma_{co} \ N_y^{co}(b, s)\   \ln
\left ( {N_y^{co}(b,
s) \over N_f} \right )\ \theta (N_y^{co}(b, s) - N_f)  \right
] \ \ \ . \label{2e}
\eeq

\noi This formula is obtained assuming longitudinal boost invariance of
hadronic densities and isoentropic longitudinal expansion (i.e. a
decrease of densities with proper time in $1/\tau$). Transverse
expansion is neglected.
$N_y^{co}(b, s)$ is the density of hadrons per unit transverse area
$d^2s$ and
per unit rapidity at impact parameter $b$. All species of hadrons are
included in
$N_y^{co}$. (If we consider only the process $\psi + \rho \to D +
\bar{D} +
\cdots$ as in  Ref. \cite{9r}, the value of $N_y^{co}(b, s)$ has to be
decreased and that
of $\sigma_{co}$ increased by the same percentage amount.)

In order to have a smooth onset of the co-movers and to avoid any
threshold effect, it is natural to take for $N_f$ the density of hadrons
per unit rapidity in a $pp$ collision, i.e. $N_f=[3/(\pi R_p^2)]\ dN^-/dy
(y^*=0) \simeq 1.15$ fm$^{-2}$. This coincides with the value introduced
in Ref. \cite{6r}. Because of this choice of $N_f$, the $\theta$-function
in Eq. (\ref{2e}) is irrelevant; we hace checked it
numerically.
Thus, $N_f$ cannot be regarded just as a free parameter.
Moreover, small changes in the value of $N_f$ can be compensated by
smaller changes in $\sigma_{co}$, without spoiling the quantitative
comparison to the data.
The argument of
the log
is the interaction time of the $J/\psi$ with co-moving hadrons. In Ref.
\protect{\cite{10r}} a different expression for the
interaction
time based on interferometry radii was used. The results obtained with
the two
expressions are practically
identical. More precisely the expression (7)
in Ref. \protect{\cite{10r}}, with the
initial
time $\tau_0 = 1$ fm/c
used there, gives practically the same results as our Eq.
(\protect{\ref{2e}}) with $N_f = 1.15$ fm$^{-2}$.

$\sigma_{co}$ is the
co-mover cross-section properly averaged over the momenta of the colliding
particles
(the relative velocity of the latter is included in its definition) and
over the different species of secondaries.
Unfortunately, the value of $\sigma_{co}$ is not known experimentally.
This is, of course, the main limitation of the co-mover scenario.
Different theoretical
calculations of the $J/\psi$-hadron cross-section based on the multipole
expansion in QCD
\cite{kaka} differ from those which include other non-perturbative effects
\cite{qua} by at
least
a factor 20 for $\sqrt{s} \sim 5$ GeV. Other references \cite{hufner}
obtain values for the $J/\psi$-$N$ cross-section at high energy of $4\div 6$ mb
and a ratio $\sigma^{\psi^\prime -N}/\sigma^{J/\psi - N} \sim 3\div 4$
in agreement with geometrical considerations. Our value
$\sigma^{\psi^\prime}_{co}/\sigma^\psi_{co}=10$
(see below) is much larger than its asymptotic value. This is consistent
with the very different behaviour of the two cross-sections near
threshold. Note, however, that Eq. (\ref{2e}) is the result of an
integration from time $\tau_0$ to freeze-out. For times
close to $\tau_0$, one is dealing with a dense interacting parton system
and thus the precise relation between $\sigma_{co}$ and the
$J/\psi(\psi^\prime)$-hadron cross-section is not established.
In this situation, inverse kinematic experiments could help to
determine the actual r$\hat{{\mbox o}}$le of co-movers in $J/\psi$
suppression.
Phenomenologically, the value of $\sigma_{co}$ obtained
here allows to make predictions at other energies, in particular for
RHIC \cite{acf}.

Note that $S^{co}(b,
s) = 1$ for
$\sigma_{co}=0$. The effects of the co-movers in proton-nucleus collisions turn
out to be negligibly small.

The inclusive cross-section for $J/\psi$ production in nuclear collisions is
then given by

\beq
I_{AB}^{\psi}(b) = \frac{I_{NN}^{\psi}}{\sigma_{pp}}
 \int d^2s \ m(b, s) \ S^{abs}(b, s) \
S^{co}(b, s) \ \ \ , \label{3e}
\eeq

\noi where

\beq
m(b, s) = AB \ \sigma_{pp} \ T_A(s) \ T_B(b -s) \ \ \ .
\label{4e}
\eeq

\noi We use $\sigma_{pp}= 30$ mb.
From Eqs. (\ref{3e}) and (\ref{4e}) we see that for Drell-Yan pair
production
($\sigma_{abs} = \sigma_{co} = 0$), $I_{AB}^{DY} = AB\ I_{NN}^{DY}$.

In the dual parton model (DPM), $N_y^{co}(b, s)$ is given by \cite{21r,22r}

\[ N_y^{co}(b, s) = \left [ N_1 \ m_A(b, s) + N_2 \ m_B(b, b-s) + N_3 \ m(b,s)
\right ]\theta ( m_B(b,b-s) - m_A(b,s))
\]
 \beq
 + [N^\prime_1 \ m_A(b, s) + N^\prime_2 \ m_B(b, b-s) + N^\prime_3 \
m(b, s) ] \theta \left ( m_A(b, s) - m_B(b, b-s) \right ) \ \ \ .
\label{5e}
\eeq

\noi Here $m$ is given by Eq. (\ref{4e}) and $m_A$, $m_B$ are the well known
geometric
factors \cite{22r,23r}

\beq
m_{A(B)}(b, s)
= A(B) \ T_{A(B)} (s) \left [ 1 - \exp \left ( - \sigma_{pp} \ B(A) \
T_{B(A)}(b - s) \right ) \right ] \ \ \ . \label{6e}
\eeq

\noi The coefficients $N_i$ and $N^\prime_i$
are obtained in DPM by convoluting momentum
distribution functions and fragmentation functions \cite{21r}. Their values
(per
unit rapidity) for the rapidity windows and energies of the NA38 and NA50
experiments are given
in Table 1. The rapidity density of hadrons is given by

\beq
{dN^{co} \over dy} = {1 \over \sigma_{AB}} \int d^2b \int d^2s
\ N_y^{co}(b, s) \
\ \ ,  \label{7e}
\eeq
with $\sigma_{AB}=\int d^2b \ (1-\exp{[-\sigma_{pp} ABT_{AB}(b)]})$,
$T_{AB}(b) = \int d^2s\ T_A(s)T_B(b-s)$.

The obtained densities
of negative hadrons
at $y^* = 0$ for $pp$, $SS$, $SAu$ and $PbPb$ are compared in Table 2
with available data \cite{24r,na35,na49},
using in each case the centrality criteria
(in percentage of total events) given by the experimentalists. At this
point we would like to comment on the differences between DPM and the
scaling in the number of participants known as wounded nucleon model
(WNM, for a review see e.g. \cite{bialas}).
The former gives a multiplicity that increases
faster with centrality. This is due to the presence of strings of type
quark-antiquark \cite{21r}. Since these strings contribute only at
midrapidity, the difference between the two models is maximal at
$y^* \sim 0$, where the NA38/NA50 spectrometer is located, and quite
small
at the negative values of $y^*$ of the NA50 $E_T$ calorimeter.
The data of the NA35 \cite{na35} and NA49 \cite{na49}
Collaborations on the rapidity distribution of negatives in central $SA$
and $PbPb$ interactions show an agreement with the WNM in the
fragmentation region and an excess of $20 \div 30$ \% at $y^* \sim 0$,
as expected from DPM.
We have checked \cite{acf} that the correlation between $E_T$ and
$E_{ZDC}$ (the energy in the NA50 zero degree calorimeter) in $PbPb$
has a small
concavity in DPM, resulting in a fit to the measured correlation which
is at least as good as the straight line obtained in the WNM.
\par \vskip 5 truemm

\noi \underbar{\bf E$_{\bf T}$ dependence:}
To determine the $E_T$ dependence we need the $E_T-b$ correlation, i.e. the
$E_T$ distribution at each impact parameter $P(E_T, b)$.
The $E_T$-dependence of the ratio $J/\psi$ over Drell-Yan is
given by

\beq
R(E_T) = {\int d^2b \ P(E_T, b) \ I_{AB}^{J/\psi} (b)
\over \int d^2b \ P(E_T, b) \
I_{AB}^{DY}(b)}\ \ \ ,  \label{8e}
\eeq

\noi where $I_{AB}^{\psi}(b)$ is given by Eq. (\ref{3e}) and $I_{AB}^{DY}(b)$
is
obtained (up to a normalization constant) from Eq.
(\ref{3e}) with $\sigma_{abs} =
\sigma_{co} = 0$.

It is clear from the discussion below Eq. (\ref{7e}) that, in the region of the
NA50 calorimeter, $P(E_T,b)$ in DPM is very similar to the WNM one
\cite{6r,26r}. However, for consistency, we are going to use the DPM
distribution:
\beq
P(E_T,b)=\frac{1}{\sqrt{2\pi q^2 a N_y^{co}(b)}}\ \exp{\left[
-\frac{[E_T-qN_y^{co}(b)]^2}{2q^2aN_y^{co}(b)}\right]}\ \ \ , \label{9e}
\eeq
where $q$ and $a$ are free parameters and $N_y^{co}(b)=\int d^2s
\ N_y^{co}(b, s)$ is obtained from (\ref{5e}) with the coeficients $N_i$
and $N^\prime_i$
obtained in DPM. They correspond to the density of neutral particles in
the rapidity windows of the NA38 ($SU$) and NA50 ($PbPb$) calorimeters.
The parameters $q$ and $a$ are obtained from a fit of the $E_T$
distributions for dimuon pair production above the $J/\psi$ mass. The
resulting fits are quite good \cite{acf} and give: $q=0.65$ GeV,
$a=1.5$ for $SU$ and $q=0.78$ GeV, $a=1.5$ for $PbPb$. It is interesting to
note that the value of $q$ for $PbPb$ is identical from the one obtained
from the best fit to the $E_T-E_{ZDC}$ correlation. The value of $a$ is
poorly determined but affects very little the results below. A value
$a=1$ \cite{6r} is also consistent with the data. Our value $a=1.5$ is
the one expected from a Poissonian distribution of clusters (resonances)
with an average cluster multiplicity of 1.5. This value agrees with the
one
obtained when clusters are identified with a realistic mixture of known
resonances and direct particles \cite{pic}.
\par \vskip 5 truemm

\noi \underbar{\bf Numerical results:}
We present the results for $J/\psi$ and $\psi^\prime$
suppression obtained with two sets of parameters.
Set I corresponds to nuclear absorption alone:
$\sigma_{abs}=7.3$ mb and $\sigma^\psi_{co}= \sigma^{\psi^\prime}_{co}
= 0$. Set II contains the effect of the co-movers: $\sigma_{abs}=6.7$
mb, $\sigma^\psi_{co} = 0.6$ mb and $\sigma^{\psi^\prime}_{co}= 6.0$
mb ($N_f=1.15$ fm$^{-2}$ as discussed previously). The
absolute normalization (corresponding to $\sigma_{pp}^\psi$ in
$(J/\psi)/AB$ and to $\sigma_{pp}^{\psi^{(\prime)}}/\sigma_{pp}^{DY}$ in
$\psi^{(\prime)}/DY$, in the acceptance of the NA38 and NA50
experiments),
is a free parameter which, for each Set,
has been determined
from a best fit to the data.

The results for $J/\psi$ suppression versus $AB$ are presented in Fig. 1.
Nuclear absorption alone, Set I,
gives a $\chi^2/dof=1.1$; although this value is quite good, the experimental
$PbPb$
point lies well below the theoretical curve for this Set (by $\sim 3$ standard
deviations). However,
Set II
gives
a satisfactory description ($\chi^2/dof=0.2$) of all points.
We also see that the effect of co-movers is much
smaller in
$SU$ than in $PbPb$.

We turn next to the $E_T$ dependence. Using Eqs. (\ref{8e})
and (\ref{9e}) we compute the ratio
$R(E_T)$ for $SU$ and $PbPb$
in the five $E_T$ intervals of the NA38 and NA50 experiments. In order to
exhibit
all the results in the same figure we plot the ratio $R$ versus $L$. This
variable is a measure of the centrality of the collision. The average value of
$L$ in each $E_T$ bin is given in the experimental papers \cite{8r}.
However, the value of $L$ is largely irrelevant since we are comparing
the measured suppression in specific $E_T$ bins with the model
calculations in the same $E_T$ bins; it only provides a
scale for the horizontal axes. For consistency, we have to take, for
each $E_T$, the same value of $L$ used by the experimentalists.
Note that the first calculations of $L$
by NA50 \cite{7r} used a sharp-surface approximation for the nuclear
density. More recent calculations \cite{8r} are based on a Saxon-Woods
density and are in better agreement with other calculations available in
the literature (e.g. \cite{vogt}).

The results for Sets I and II for $J/\psi$ suppression ($(J/\psi)/DY$)
are
given in Fig. 2 for all $pA$ and $AB$ data as a function of $L$. In
Fig. 3 the same results for $SU$ and $PbPb$ are presented
as a function of $E_T$ in the form of a continuous line (not as the
average in each $E_T$ bin as in Figs. 2 and 4).
Set I gives a $\chi^2/dof =27.9$ for all $pA$, $SU$ and
$PbPb$ data,
indicating that nuclear absorption
alone fails very badly. On the contrary, without the $PbPb$ data the
best fit with Set I gives $\chi^2/dof =0.9$.
Set II gives $\chi^2/dof = 2.7$ with only $pA$ and $SU$ data. So without
the $PbPb$ data it is hardly possible to decide whether co-movers are present
or not -- although the $\chi^2/dof$ is better without co-movers.
When $PbPb$ data are included Set II gives $\chi^2/dof = 4.3$.
What prevents this value of being smaller 
is the peculiar $L$ or $E_T$ shape of the
$PbPb$ data which cannot be reproduced in our simple approach.

The results for $\psi^\prime$ suppression ($\psi^\prime/DY$)
are presented in Fig. 4. Set
I gives  $\chi^2/dof =$ 14.3 and Set II  $\chi^2/dof =$ 1.3 for all
$pA$, $SU$ and $PbPb$ data.

The treatment of the co-movers presented above is similar to the one in
Ref. \cite{6r}. However, it differs from it and from previous
treatments \cite{9r,10r} in that we use the
DPM expression for the density of hadrons,
Eq. (\ref{5e}), instead of assuming it to be proportional to either the
number of participants or to $E_T$. Another difference resides in the
nuclear densities. In the present work
(and also in \cite{12r}) calculations have been done with the
nuclear density described after Eq. (\ref{1e}), whereas in
\cite{6r,vqgp} the 3-parameter Fermi distribution of Ref. \cite{jvv}
is used. Using the latter and keeping all other parameters as above, we
have obtained a $J/\psi$ suppression between the first and the last
$E_T$ bins which is 7 \% larger in $SU$ and 4 \% larger in $PbPb$.

In conclusion, the data on both $J/\psi$ and $\psi^\prime$ suppression
can be described in a co-mover approach with a small number of free
parameters, which take reasonable values. In this approach there is no
discontinuity in any observable. One obtains a monotonic decrease of the
$J/\psi$ and $\psi^\prime$ over Drell-Yan ratios from the most
peripheral to the most central collisions. A clear departure from such a
behaviour would rule out the co-mover description of $J/\psi$ and
$\psi^\prime$ suppression presented above. It is also important to
compute the $J/\psi$ suppression at RHIC in the two approaches using the
parameters determined at SPS. This suppression is expected to be quite
large and will possibly be wildly different in the two scenarios.

\noindent {\bf Acknowledgments:}
We thank E. G. Ferreiro, C. Gerschel, A. Krzywicki, C. Louren\c co,
C. A. Salgado and J. Tr$\hat{{\mbox a}}$n
Thanh V$\hat{{\mbox a}}$n for useful
discussions, J.-P. Blaizot and J.-Y. Ollitrault for discussions and for
providing us with
their numerical code to cross-check our own, M. Ga\'zdzicki,
C. Gerschel, L. Kluberg
and C. Louren\c co for information on
experimental data, and D. Kharzeev, M. Nardi and H. Satz, who pointed
out
to us a numerical error in the first version of this work.
Also N. A. thanks the Xunta de Galicia for financial
support.
Laboratoire de Physique Th\'eorique et Hautes Energies is
laboratoire associ\'e au Centre National de la Recherche
Scientifique - URA D0063.

\noindent {\bf Note added}: A mistake in the evaluation of the
experimental
errors of both $SU$ and $PbPb$ data has been reported by A. Romana (NA50
Collaboration) at the XXXIIIrd Rencontres de Moriond (Les Arcs, France,
March 1998). The statistical errors of $SU$ ($PbPb$) have to be
multiplied by a factor 3 (1.4). This reduces significantly our
$\chi^2/dof$ for ($(J/\psi)/DY$), which, for Set I (II), are now 0.6
(1.4)
for $pA$ and $SU$ and 8.1 (1.9)
for all systems ($pA$, $SU$ and $PbPb$) together.

\newpage
\centerline{\bf Table captions:}
\vspace{1cm}

\noi {\bf Table 1.} Coefficients
(per unit rapidity)
in Eq. (\ref{5e}) in the rapidity
windows of the electromagnetic calorimeters (corresponding to neutral
multiplicity) and dimuon detector (corresponding to multiplicity of
charged plus neutrals)
and at energies of the NA38 and NA50
experiments. When using these
values in Eq. (\ref{5e}) one should put $A=A_{projectile}$ and
$B=A_{target}$.

\noi {\bf Table 2.} Negative particle densities at midrapidity obtained
with Eqs. (\ref{5e}) and (\ref{7e}) ($th$),
compared to experimental data \cite{24r,na35,na49}
($exp$). Percentages of total events
(given in the experimental papers), and corresponding
impact parameters considered, are shown.

\newpage
\centerline{\bf Figure captions:}
\vspace{1cm}

\noi {\bf Figure 1.} $J/\psi$ suppression versus $AB$: Set I (dotted
line) and Set II (solid line) compared to the
experimental data \cite{8r}. The normalization factors
are 2.01 nb/nucleon$^2$ for Set I and 2.08 nb/nucleon$^2$ for Set II.
Note that the calculations have been
performed
only for those nuclei where data exist. The obtained values have
been
joined by straight lines.

\noi {\bf Figure 2.} Ratio $(J/\psi)/DY$
(Eq. (\ref{8e})) versus $L$ (fm): Set I
(dotted
line) and Set II (solid line)
compared to the
experimental data \cite{8r}. Results obtained as an average over each
experimental $E_T$ bin have been joined by straight lines.
The normalization factors
for the
theoretical lines (giving the $\chi^2/dof$ indicated in the text for all
$pA$, $SU$ and $PbPb$ data included in the fit)
are 38.32 for Set I and 45.50 for Set II.

\noi {\bf Figure 3.} Ratio $(J/\psi)/DY$
(Eq. (\ref{8e})) versus $E_T$ (GeV) compared to the
experimental data \cite{8r},
in the form of a continuous line (not as an
average over each $E_T$ bin as in Figs. 2 and 4), for $SU$ (upper figure)
and $PbPb$ (lower figure). Conventions and
normalizations are the same as in Fig. 2.

\noi {\bf Figure 4.} Ratio $\psi^\prime/DY$
(Eq. (\ref{8e})) versus $L$ (fm): Set I
(dotted
line) and Set II (solid line)
compared to the
experimental data \cite{8r}. Results obtained as an average over each
experimental $E_T$ bin have been joined by straight lines.
The normalization factors
for the
theoretical lines (giving the $\chi^2/dof$ indicated in the text for all
$pA$, $SU$ and $PbPb$ data included in the fit)
are 0.299 for Set I and 0.723 for Set II.

\newpage

\centerline{   }
\vspace {2cm}
\centerline{\bf Table 1}
\vspace{1cm}
\begin{center}
\begin{tabular}{ccccccc} \hline \hline
 & $N_1$ & $N_2$ & $N_3$ & $N_1^\prime$ & $N_2^\prime$ & $N_3^\prime$
\\ \hline
$SU$ $(-1.2<y^*<1.2)$ & 0.2096 & 0.2746 & 0.1598 & 0.2827 & 0.2015 &
0.1598 \\
$PbPb$ $(-1.8<y^*<-0.6)$ & 0.3549 & 0.0548 & 0.0946 & 0.3198 &
0.0899 & 0.0946 \\
$SU$ $(0.0<y^*<1.0)$ & 0.8433 & 0.6003 & 0.4995 & 1.0854 & 0.3582
& 0.4995 \\
$PbPb$ $(0.0<y^*<1.0)$ & 0.5891 & 0.8086 & 0.4248 & 0.3685 & 1.0292
& 0.4248 \\
\hline \hline
\end{tabular}
\end{center}

\vspace {6cm}
\centerline{\bf Table 2}
\vspace{1cm}
\begin{center}
\begin{tabular}{ccc} \hline \hline
Reaction & $dN^-/dy|_{y^*=0}^{th}$ & $dN^-/dy|_{y^*=0}^{exp}$ \\ \hline
$pp$ & 0.73 & $0.76\pm 0.04$\\
$SS$ (11 \%, $b \leq 2.7$ fm) & 19.3 & $19.0\pm 1.5$\\
$SAu$ (1.3 \%, $b \leq 1.3$ fm) & 56.3 & $59.0\pm 3.0$\\
$PbPb$ (5 \%, $b \leq 3.4$ fm) & 207 & $195 \pm 15$\\
\hline \hline
\end{tabular}
\end{center}

\newpage

\centerline{\bf Figure 1}
\vspace{1cm}

\epsfig{file=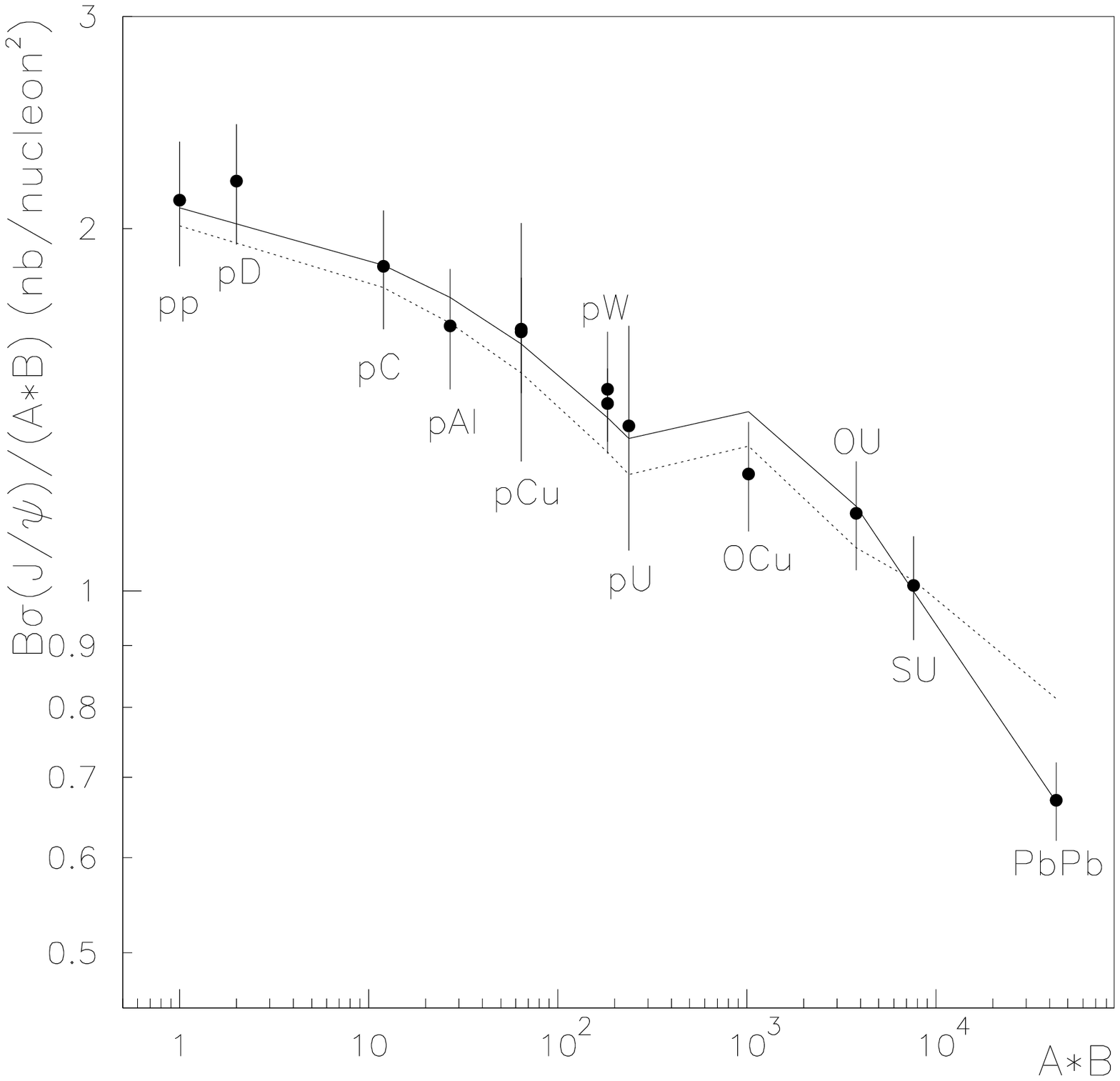,width=15.25cm}

\newpage

\centerline{\bf Figure 2}
\vspace{1cm}

\epsfig{file=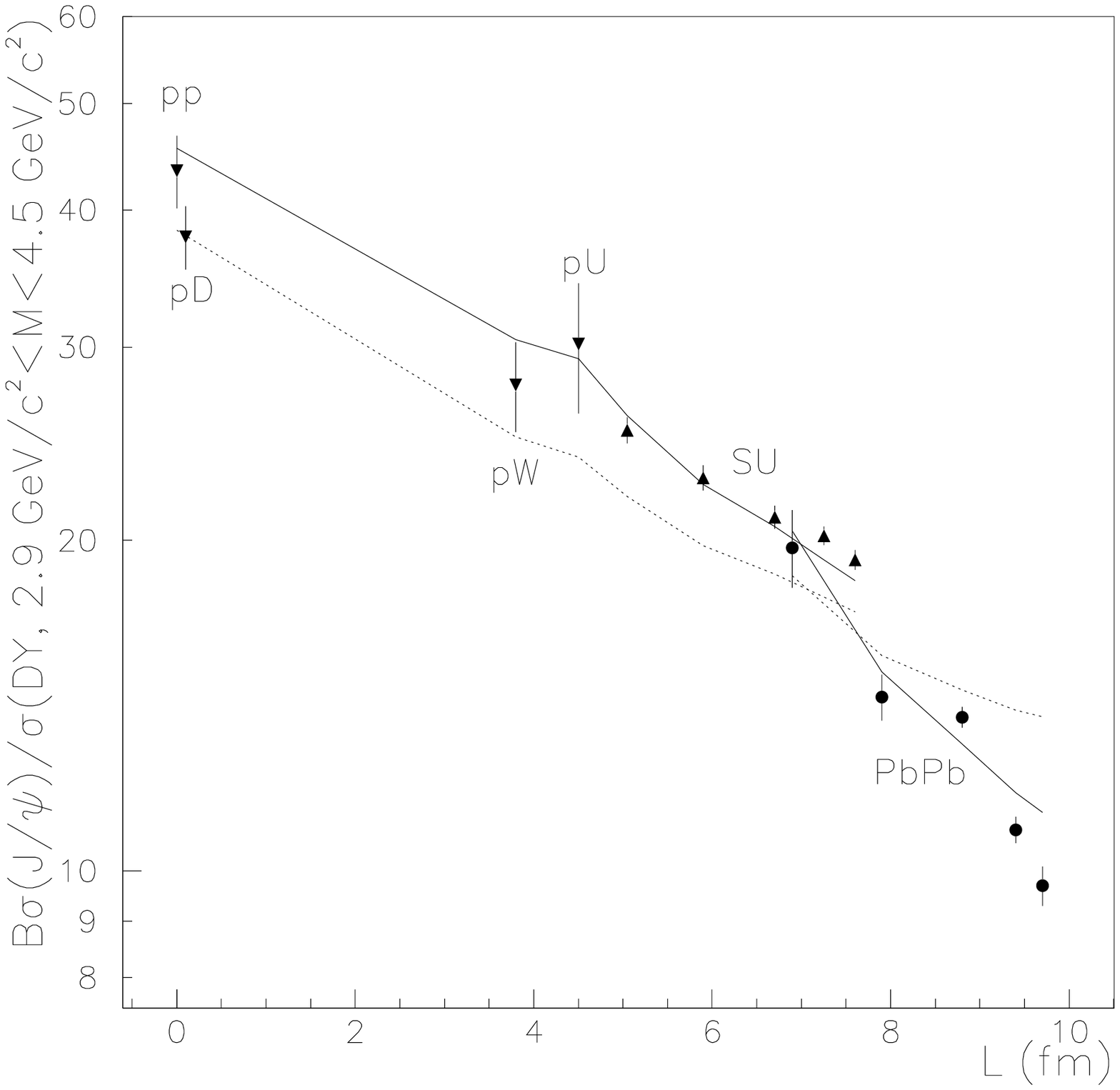,width=15.25cm}

\newpage

\centerline{\bf Figure 3}
\vspace{1cm}

\epsfig{file=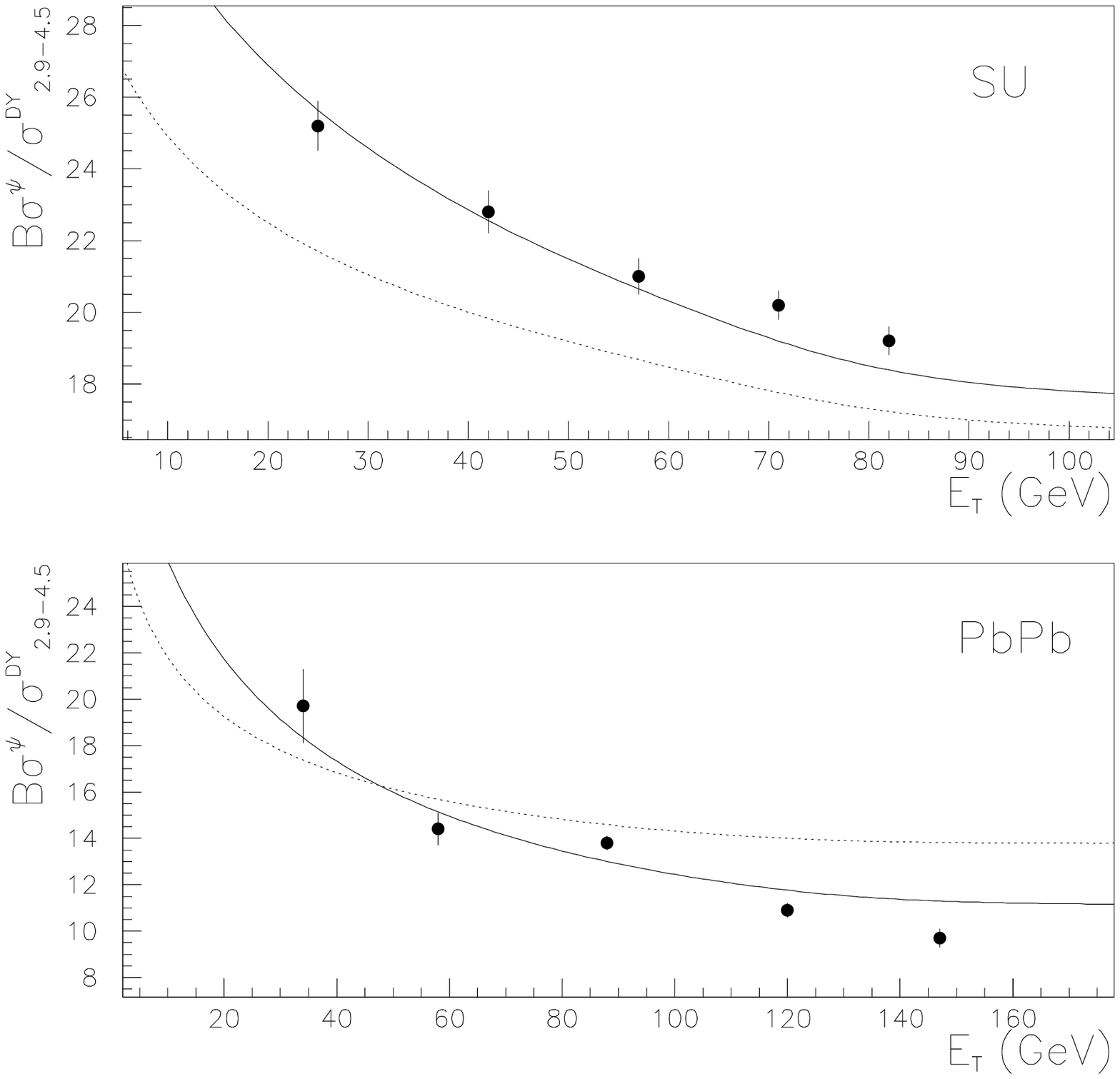,width=15.25cm}

\newpage

\centerline{\bf Figure 4}
\vspace{1cm}

\epsfig{file=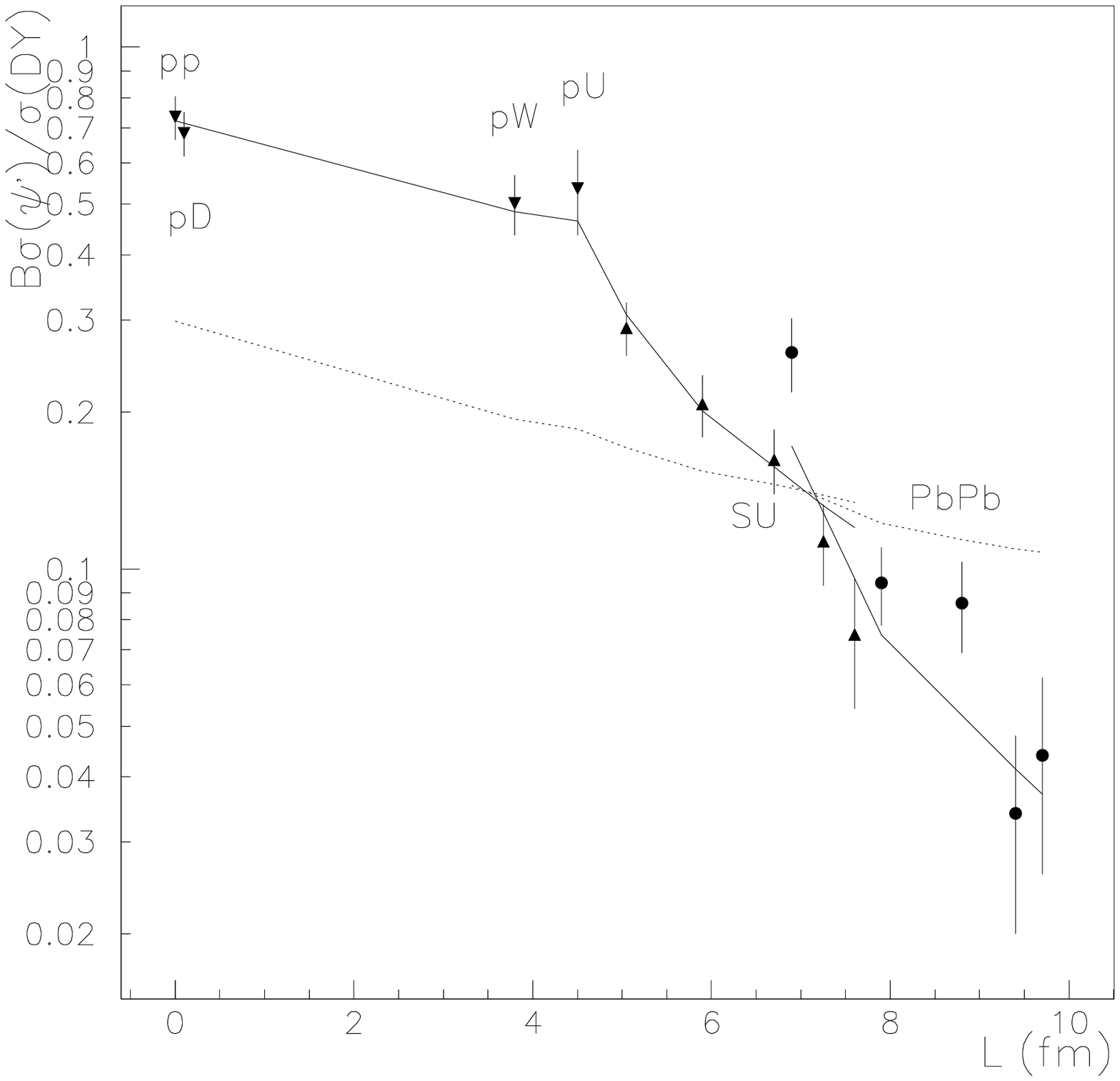,width=15.25cm}

\end{document}